%% file: SOC.tex
\documentclass[12pt]{article}

\RequirePackage[OT1]{fontenc}
\usepackage{amsthm,amsmath,natbib}
\usepackage[]{authblk}
\usepackage{float}
\usepackage{lipsum}
\usepackage{lineno}
\usepackage[margin=3cm]{geometry}
\usepackage[hyphenbreaks]{breakurl}
\usepackage[hyphens]{url}
\RequirePackage[colorlinks,citecolor=blue,urlcolor=blue]{hyperref}
\input{mylatex.tex}

\input{packages.tex}

\title{Modeling soil organic carbon with Quantile Regression: Dissecting predictors' effects on carbon stocks}

\author
{Luigi Lombardo$^1$$^,$$^2$$^*$, Sergio Saia$^3$, Calogero Schillaci$^4$, P. Martin Mai$^2$, Rapha\"el Huser$^1$\ \\

\normalsize{$^{1}$ Computer, Electrical and Mathematical Sciences \& Engineering Division,
KAUST, Thuwal, Saudi Arabia}\\
\normalsize{$^{2}$ Physical Sciences and Engineering Division, KAUST, Thuwal, Saudi Arabia}\\
\normalsize{$^{3}$Council for Agricultural Research and Economics (CREA) \\ Cereal and Industrial Crops Research Centre (CREA-CI), Foggia, Italy}\\
\normalsize{$^{4}$Department of Agricultural and Environmental Science, University of Milan, Italy}
}
\date{\today}
\begin{document}
	\maketitle
 \begin{center}
{\large{\bf Abstract }} 
\end{center}
Soil Organic Carbon (SOC) estimation is crucial to manage both natural and anthropic ecosystems and has recently been put under the magnifying glass after the Paris agreement 2016 due to its relationship with greenhouse gas. Statistical applications have dominated the SOC stock mapping at regional scale so far. However, the community has hardly ever attempted to implement Quantile Regression (QR) to spatially predict the SOC distribution. In this contribution, we test QR to estimate SOC stock (0-30 $cm$ depth) in the agricultural areas of a highly variable semi-arid region (Sicily, Italy, around 25,000 $km2$) by using topographic and remotely sensed predictors. We also compare the results with those from available SOC stock measurement. The QR models produced robust performances and allowed to recognize dominant effects among the predictors with respect to the considered quantile. This information, currently lacking, suggests that QR can discern predictor influences on SOC stock at specific sub-domains of each predictors. In this work, the predictive map generated at the median shows lower errors than those of the Joint Research Centre and International Soil Reference, and Information Centre benchmarks. The results suggest the use of QR as a comprehensive and effective method to map SOC using legacy data in agro-ecosystems. The R code scripted in this study for QR is included.\\  

{\bf Keywords:} Quantile Regression, R coding, Topsoil Organic Carbon, Digital Soil Mapping, Mediterranean agro-ecosystem\\

{\bf Corresponding Author:} Luigi Lombardo*, Email: luigi.lombardo83@gmail.com

\section{Introduction}
Soil Organic Carbon (SOC) plays a key role in various agricultural and ecological processes related to soil fertility, carbon cycle and soil-atmosphere interactions including CO$_2$ sequestration. Thus, its knowledge has a crucial importance both at global and local scales, especially when aiming at managing natural, anthropic areas and especially agricultural lands. In this context, the scientific community has spent considerable efforts in mapping SOC, modeling its spatiotemporal variation and confirming its primary role in shaping ecosystems functioning \citep{ajami2016environmental,grinand2017estimating,ratnayake2014changes,schillaci2017spatio}. 

Spatiotemporal studies can be found in various geographic contexts from Africa \citep{akpa2016total}, Asia \citep{chen2016impacts}, Australia \citep{henderson2005australia}, Europe \citep{yigini2016assessment}, North-America \citep{west2002modeling} to South-America \citep{araujo2016assessment}. The variability of the local landscape, available funding, mean gross income of the population in the area and temporal commitment affect the number of samples, their spatial density and distribution. As a result, there are experiments conducted on almost regular and dense grids, most of which focus on small areas \citep{lacoste2014high,taghizadeh2016digital} and other where the sampling strategy significantly varies across space \citep{mondal2016impact}. The latter studies mainly correspond to regional or even greater scales \citep{reijneveld2009soil,sreenivas2016digital}, with only few cases where an optimal sample density is maintained at a national level \citep{mulder2016national}. The characteristics of environment under study can require the use of different predictors capable of explaining the variability of soil traits, topography and standing biocoenosis, especially (cropped or natural) phytocoenosis, the latter being efficiently explained by remotely sensed (RS) properties \citep{morellos2016machine,peng2015modeling}. 

Modeling procedures for SOC primarily aims at constructing present, past or predictive maps and studying the role of each predictor over the target variable. Regarding the latter, the estimation of predictor contributions on a target variable such as SOC, is of particular interest to efficiently obtain agro-environmental and social benefits \citep[e.g.][]{rossel2016soil}.

Statistical applications provide quantitative ways to deal with such research questions. The current literature encompasses algorithms that can be clustered into interpolative and predictive. Pure interpolators are broadly used when the density of the samples is sufficient to regularly describe the variation of SOC across a given area. Examples can be found \citep{hoffmann2014assessing,piccini2014estimation} with excellent performances reported. The weakness of these approaches becomes evident when using data sets with non-regular distribution in space \citep{dai2014spatial,miller2016towards}. Conversely, regression-based predictive models hardly suffer from the spatial sampling scheme as they do not rely on the distribution across the geographic space in order to derive functional relations between SOC and dependent variables \citep{hobley2016environmental}.

Among these, linear regression models are a well-established tool for estimating how, on average, certain environmental properties affect SOC and SOC stock \citep{rodriguez2015modelling}. However, they are bounded by definition to model the conditional mean, thus being unable to explore the effects of the same properties at different C contents or stock of the soil, especially at the boundaries of the distribution.

In the present work, Quantile Regression (hereafter QR, \cite{koenker2005quantile}) is used to model SOC stock from a non-homogenously sampled topsoil SOC dataset using soil texture, land use, topographic and remotely sensed covariates. In particular, QR is able to model the relationship between a set of covariates and specific percentiles of SOC. In classical regression approaches, the regression coefficients (also often called beta coefficients) represent the mean increase in the response variable produced by one unit increase in the associated covariates. Conversely, the beta coefficients obtained from QR represent the change in a specific quantile of the response variable produced by a one unit increase in the associated covariates. In this way, QR allows one to study how certain covariates affect, for example SOC median (quantile $\tau=0.5$) or extremely low (e.g., $\tau=0.05$) or high (e.g., $\tau=0.95$) SOC values. 
Therefore, it gives a much more complete description of the effect of predictors on the whole SOC probability distribution (i.e., not just the mean) and thus offer the chance to study differential SOC responses to environmental factors.

Furthermore, when used for mapping purposes, QR also allows for soil mapping at given quantiles, providing analogous estimates to more common approaches by using the median instead of the mean.

In the present experiment we use a nested strategy to model SOC in Sicilian agricultural areas with QR: we initially aim at testing the QR overall performances when modeling the SOC stock by segmenting its distribution into 19 quantiles ($\tau=0.05$ to $\tau=0.95$). Subsequently, we examine the coefficients of each predictor for each of the quantiles. Ultimately, we compare the median prediction with available SOC benchmarks for the same study area to test the efficiency of QR for soil mapping purposes. The dataset used in this contribution is the same used in \cite{schillaci2017modelling} where a Stochastic Gradient Treeboost is adopted. 

\section{Materials and methods}
\label{sec:Materials and methods}

\subsection{Study area}
\label{sec:Study area}

Sicily with its approximate 25 thousand squared kilometers is the biggest Mediterranean island. More than 60\% of its area is cropped.~The natural/semi natural ecosystems include i) Mediterranean maquis, ii) dunes and coastal systems, iii) woods and forests. There are also 37 ancillary islands that are not considered in the present study. Sicily has several sub-climatic zones, all of which are included in hot-summer Mediterranean climate (Csa Koeppen) and warm-summer Mediterranean climate (Csb Koeppen) with mean annual temperatures usually higher than 15.8$^\circ$ C. From the West to the South-East coasts, indicators of a semiarid environment can be observed over the year with low or no rainfall summer, high air temperatures and evapo-transpiration demand together with water deficit. The mountainous areas (Madonie, Sicani, Nebrodi and Peloritani ridges, physiography can be checked in \cite{schillaci2017spatio}) are scarcely cultivated mostly because of conservation policies acting in favor of the local temperate woodland. The continentality index, which is determined by the difference between the mean air temperature during summer and winter, is similar in all the climatic subregions.

According to the latest soil map published by \cite{fantappie2010factors} using the World Reference Based \citep{iuss2014world}  soil classification, the dominant soils in Sicily are: Entisols (36\%), Inceptisols (34\%), followed by the Mollisols, Alfisols, Vertisols and Andisols. This climatic context plays an important role on the decay processes of organic residue \citep{lutzow2006stabilization} and on the stabilization of organic fractions. In particular, the local climatic setting facilitates the decomposition and mineralization of the organic matter.

\subsection{SOC Data}
\label{sec:SOC Data}

\begin{figure}[t!]
\centering
\includegraphics[width=\linewidth]{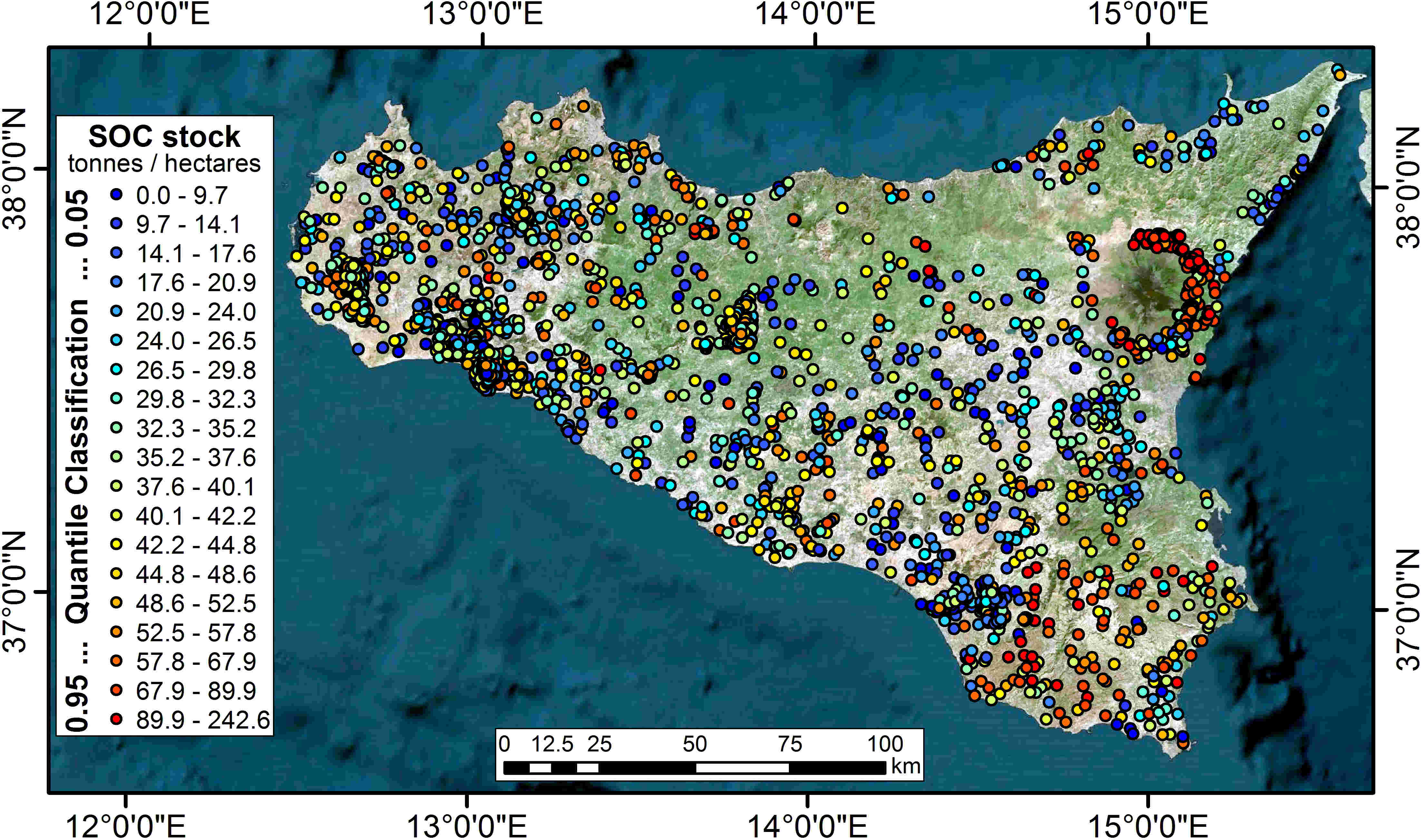}
\caption{SOC stock dataset and geographic contextualization.}
\label{fig:SOCgeog}
\end{figure}

The available datasets represent the SOC stock (expressed in $ton {\cdot} ha^{-1}$) of the topsoils (Ap horizon, from 0 to 30$cm$ depth) primarily from agricultural areas (Figure \ref{fig:SOCgeog}). It has been calculated from the organic carbon (expressed in $g {\cdot} kg^{-1}$) multiplied by the soil bulk density. The latter is derived by pedotransfer function \citep{pellegrini2007new}. In total, 2202 samples are used for modeling purposes. See \citep{schillaci2017modelling} for further information on the dataset.

Supplementary Figure 1 
shows the variability associated with each of the considered quantiles. The dataset was provided by the Assessorato Regionale Territorio Ambiente (ARTA) as georeferenced SOC values derived by pedological profiles.

The adopted covariates and their interpretation are discussed in the Supplementary Materials, Predictors Section. 
The distribution of the aforementioned covariates is shown in Supplementary Figure 2 
through their Empirical Cumulative Distribution Function.
Prior to any analysis, we transformed some of the variables.~This is shown and explained in the Supplementary Material (Figure 3 and Pre-processing Section, respectively).


\subsection{Statistical modeling using quantile regression}
\label{sec:QR}
\subsubsection{Quantile regression}

In classical regression analysis, the fluctuations in the mean of a response variable (e.g., $\log(\mbox{SOC})$) are typically explained through a linear function of a set of predictors. In the case where $n$ responses $Y_1,\ldots,Y_n$ are observed with their $p$ respective predictors $x_{1i},\ldots,x_{pi}$ (here assumed to be continuous for simplicity), a statistical model may be formulated as
\begin{equation*}\label{eq:classicalreg}
Y_i=\beta_0+\beta_1x_{1i}+\cdots+\beta_px_{pi}+\varepsilon_i,
\end{equation*}
where the random variables $\varepsilon_i$ are typically assumed to be mutually independent and to follow a normal distribution with zero mean and finite variance $\sigma^2$. Under such a model, and if the predictors are linearly independent, the vector of unknown regression parameters $\beta=(\beta_1,\ldots,\beta_p)^T$ may be estimated using the Ordinary Least Squares (OLS) estimator $\hat\beta_(OLS)$, which may also be seen as minimizing the squared loss function, i.e.,
\begin{equation}\label{eq:classicalregbeta}
\hat\beta_{OLS}=(X^TX)^{-1}X^TY=\min_{\beta}\|Y-X\beta\|^2=\min_\beta\sum_{i=1}^n(Y_i-\beta_0-\beta_1x_{1i}-\cdots-\beta_px_{pi})^2,
\end{equation}
where $Y=(Y_1,\ldots,Y_n)^T$ is the vector of observations, and $X$ is the $n$-by-$(p+1)$ design matrix, where the first column corresponds to the intercept and is a vector of ones, and each other column corresponds to a specific predictor, i.e., it contains the values $x_{k1},\ldots,x_{kn}$, $k=1,\ldots,p$. From the right-hand side of \eqref{eq:classicalregbeta}, the conditional \emph{mean} of $Y$ may be estimated by $\hat\beta_{0;OLS}+\hat\beta_{1;OLS}x_1+\cdots+\hat\beta_{p;OLS}x_p$. In other words, this is a \emph{point predictor}, focusing on a single feature (i.e., the mean) of the distribution of the response $Y$.

More detailed information on the whole conditional (not necessarily Gaussian) \emph{distribution} of the response $Y$ may be obtained using \emph{quantile} regression. By definition, for each probability $0\leq \tau\leq1$, the $\tau$-quantile $y_\tau$ of $Y$ is the value exceeding $(100\times \tau)\%$ of the data. Mathematically, one has $\pr(Y\leq y_\tau)=\tau$, and the collection of all quantiles $\{y_\tau: 0\leq \tau\leq 1\}$ fully characterizes the probability distribution of $Y$. The value $\tau=0.5$ corresponds to the \emph{median}, while low and high quantiles (for low and high values of $\tau$, respectively) correspond to extreme values of $Y$ lying in the lower and upper tails of the distribution, respectively. 

By analogy with \eqref{eq:classicalregbeta}, the conditional $\tau$-quantile may be estimated by minimizing an objective function, where the squared loss function is replaced by the quantile loss function. More precisely, computing
\begin{equation}\label{eq:quantileregbeta}
\hat\beta_{\tau}=\min_\beta\sum_{i=1}^nL_\tau(Y_i-\beta_0-\beta_1x_{1i}-\cdots-\beta_px_{pi}),
\end{equation}
where the quantile loss function $L_\tau$ is defined as
$$L_\tau(x)=\begin{cases}
-2(1-\tau)x,&x<0;\\
2\tau x,& x\geq 0,
\end{cases}$$
the conditional $\tau$-quantile $y_\tau$ may then be estimated as $\hat y_\tau=\hat\beta_{0;\tau}+\hat\beta_{1;\tau}x_{1}+\cdots+\hat\beta_{p;\tau}x_{p}$. When $\tau=0.5$, $L_{0.5}=|x|$ is the absolute loss function, and $\hat y_{0.5}$ corresponds to the estimated conditional median. In our application, we choose a sequence of $19$ equispaced probabilities $\tau=0.05,0.1,\ldots,0.95$ to fit separate quantile regression models, giving much deeper insight into the complete conditional distribution of the SOC values, as a function of spatial predictors. By focusing on low (respectively high) quantiles, regression coefficients inform us about the predictors mostly influencing the absence (respectively high concentrations) of SOC stock over space. By considering independent quantile regression models for different values of $\tau$, this allows for the possibility that the importance of certain predictors may change according the SOC level. More statistical details on quantile regression and its application may be found in \cite{koenker2005quantile}.

Finding the estimated parameters $\hat\beta_{\tau}$ by optimizing \eqref{eq:quantileregbeta} is not trivial, but robust algorithms have been implemented and made freely available in the {\tt R} package {\tt quantreg}. Model checking and validation may be performed using classical regression techniques with some minor adjustments. For example, to assess the goodness of fit, the coefficient of determination $R^2$ is typically replaced by a similar measure based on the quantile loss, although the interpretation remains essentially the same. Similarly, to check the ability of the model to predict unobserved values, cross-validation combined with the quantile loss function is typically used, in order to be consistent with the fitting procedure, instead of using the mean squared error as in classical regression analysis.

\subsubsection{Model building strategy, estimation and uncertainty assessment}
The strategy adopted in the present work includes five steps: 

\begin{enumerate}
\item We perform a preliminary multicollinearity analysis to exclude highly correlated covariates. When Pearson's  correlation coefficients are above 0.7 or below -0.7, we remove one of two or more collinear covariates as suggested by \citep{pengelly2001}. This is shown and explained in the Supplementary Material (Figure 4 and Pre-processing Section, respectively).
\item Categorical covariates are converted into dummy variables equivalent to each predictor level. Then, the most and least representative dummy classes are removed to avoid using a singular design matrix and subsequent parameter estimates.
The least represented classes contain one to five SOC stock samples. This allows to remove potential sources of noise in the modeling procedure, whereas the effect of the most frequent class are carried in the model intercept. The most frequent classes account for a significant part of the data by definition, thus the interpretation of their contribution to the model is clearly important. To investigate their effects on SOC stock we pre-run a separate simpler model built only with the most frequent class within the covariates.

\item Model performances or predictive power is evaluated through leave-one-out cross-validation \citep{ref1}. This allows for producing quality metrics based on quantile loss \citep{koenker1978regression}. In a QR framework, the latter is equivalent to the $R^2$ coefficient used in classical linear regression.  

\item Model uncertainty over replicates is implemented through non-parametric case-resampling bootstrap \citep{davison1997bootstrap}. In particular, 10000 replicates are generated by resampling each of the 2202 cases with replacement. As a result, 10000 replicates of the beta coefficient estimates for each predictor and categorical class are produced for each of the 19 quantiles considered in this study. Similarly, 19 sets of 10000 predictive maps are also computed. This procedure evaluates the variability of the modeling output and the reliability of the final estimates across replicates. 
\item SOC regionalization is conducted by producing 19 distinct quantile predictive maps by using the original dataset without any resampling scheme to ensure the full predictive power for mapping purposes.
\end{enumerate}

\subsection{Currently available SOC estimations in the study area}
\label{sec:Benchmarks}

Three digital soil mapping products are currently available for the area under study: i) the ISRIC World Soil Information (\url{http://www.isric.org}, \cite{hengl2014soilgrids1km}), ii) the Global Soil Organic Carbon Estimates of the Harmonized World Soil Database (\url{http://esdac.jrc.ec.europa.eu/content/global-soil-organic-carbon-estimates}, \cite{hiederer2011global}) and iii) the European Joint Research Centre JRC European SOC map \citep{lugato2014new}. These layers represent the state of the art of digital soil mapping and are de facto the only SOC benchmarks for the globe and for Europe. According to \cite{hengl2014soilgrids1km}, SOC distribution is calculated through Generalized Linear Models at a 1-km resolution using the GSIF package in R. \cite{hiederer2011global} use analogous linear regression model and spatial resolution to regionalize the SOC data over the globe. Conversely, the JRC European estimates are calculated using a deterministic approach using the agro-ecosystem SOC model CENTURY \citep{parton1988dynamics}. The inclusion of such estimates in the present contribution allows to compare the regional QR prediction to reliable, robust and well tested analogous datasets. The comparison is based on the median QR prediction together with the aforementioned benchmarks. To accommodate for differences in the spatial resolution we downscale all maps to the minimum common resolution (1-km cell size) where the resulting values per pixel represent the average SOC stock among smaller pixels in a given 1-km cell side.

\section{Results}

\begin{figure}[t!]
\centering
\includegraphics[width=0.5\textwidth]{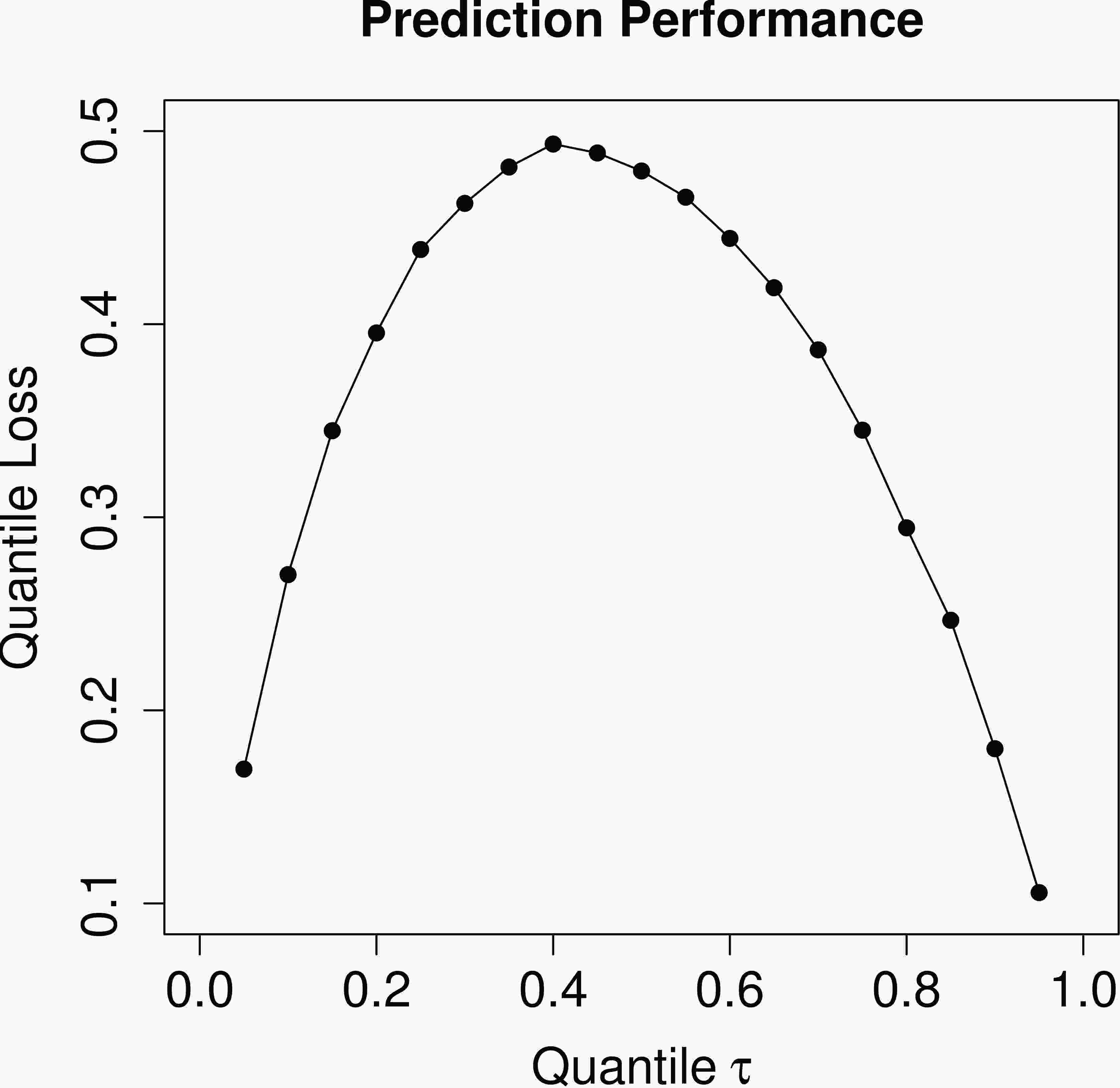}
\caption{Leave-one-out performance evaluation through quantile loss.}
\label{fig:Performances}
\end{figure}

Leave-one-out cross-validation performances appear in line with other methods in the literature. In particular, \cite{schillaci2017modelling} report an R$^2$ of 0.47 whereas the quantile loss reaches 0.49 for quantiles $\tau=0.4,0.45$ (see Figure \ref{fig:Performances}). In addition, Figure \ref{fig:Performances} reveals that the quantile loss has a bell shape as a function of the quantile level. This implies that the predictive power decreases towards the boundaries of the distribution.  

\begin{figure}[t!]
\centering
\includegraphics[width=0.7\textwidth]{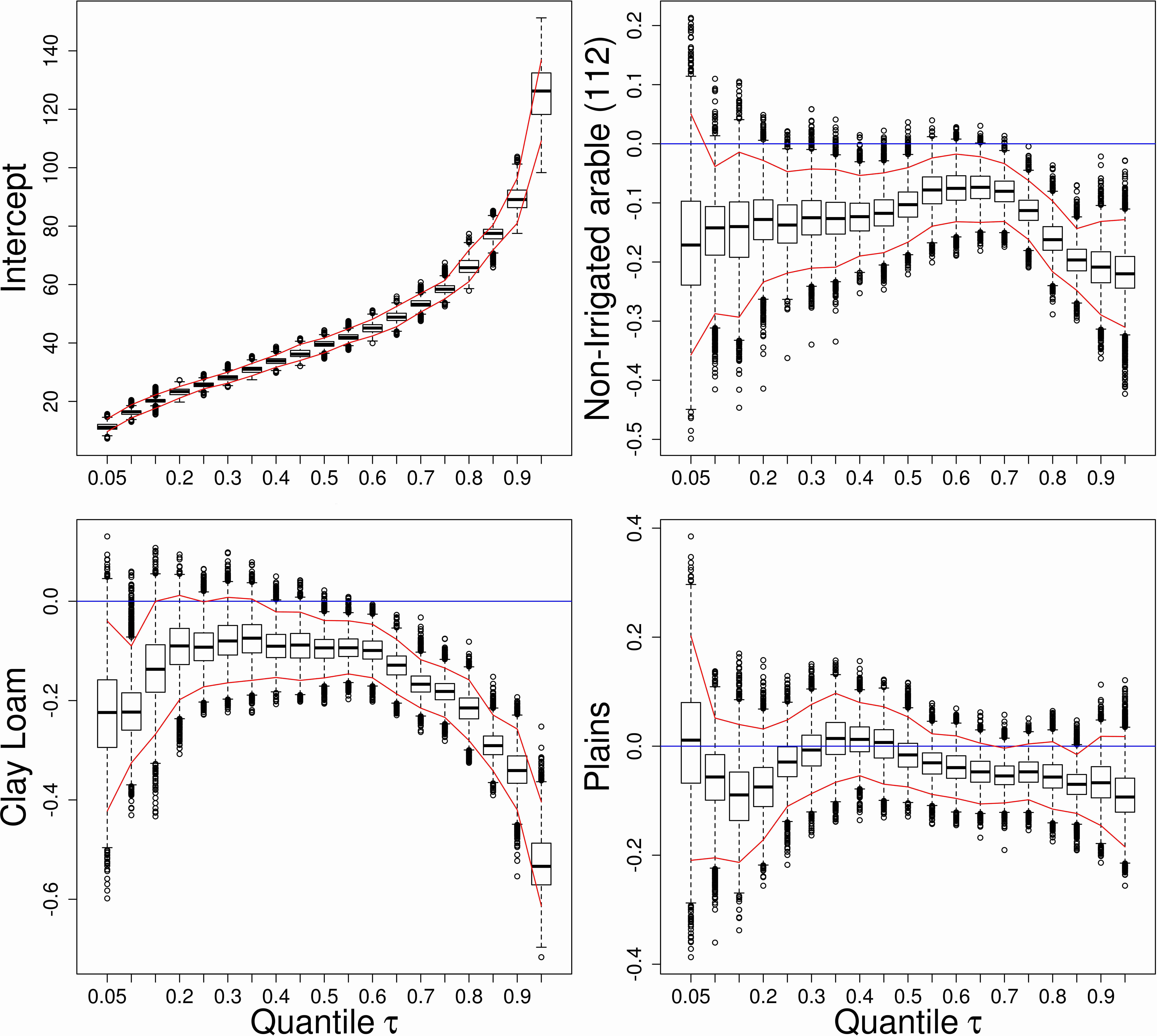}
\caption{Boxplots of estimated beta coefficients based on the simple model with 10000 bootstrap replicates, and plotted with respect to the quantile level $\tau=0.05,\ldots,0.95$. The blue line represents 0 (i.e., no effect), while the red curves are $95\%$ pointwise confidence intervals.}.
\label{fig:Beta simple}
\end{figure}

\begin{figure}[t!]
\centering
\includegraphics[width=1\textwidth]{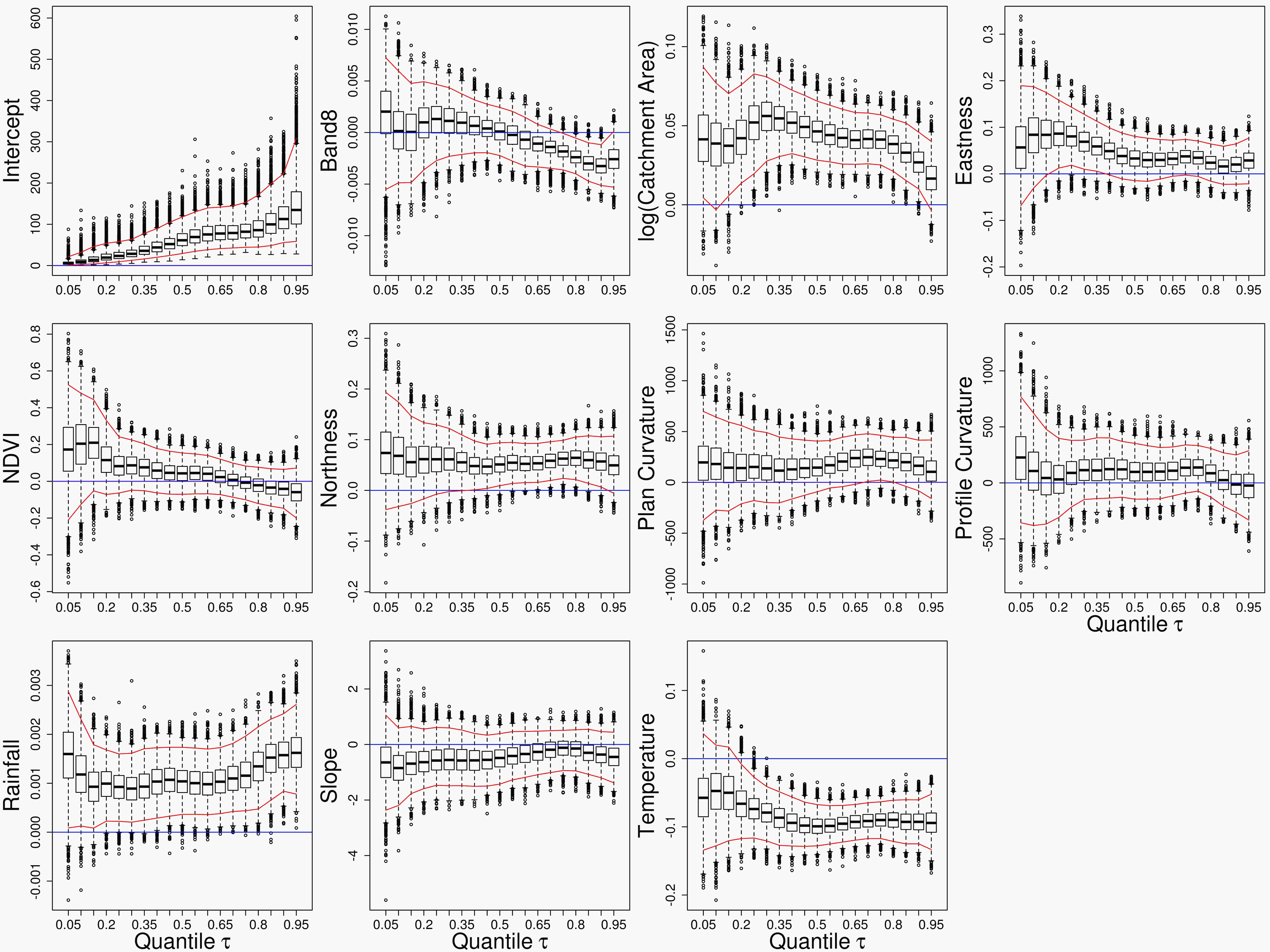}
\caption{Boxplots of estimated beta coefficients for continuous predictors. These results are based on the final model with 10000 bootstrap replicates, and plotted with respect to the quantile level $\tau=0.05,\ldots,0.95$. The blue line represents 0 (i.e., no effect), while the red curves are $95\%$ pointwise confidence intervals.}.
\label{fig:Beta Continuous}
\end{figure}

\begin{figure}[t!]
\centering
\includegraphics[width=1\textwidth]{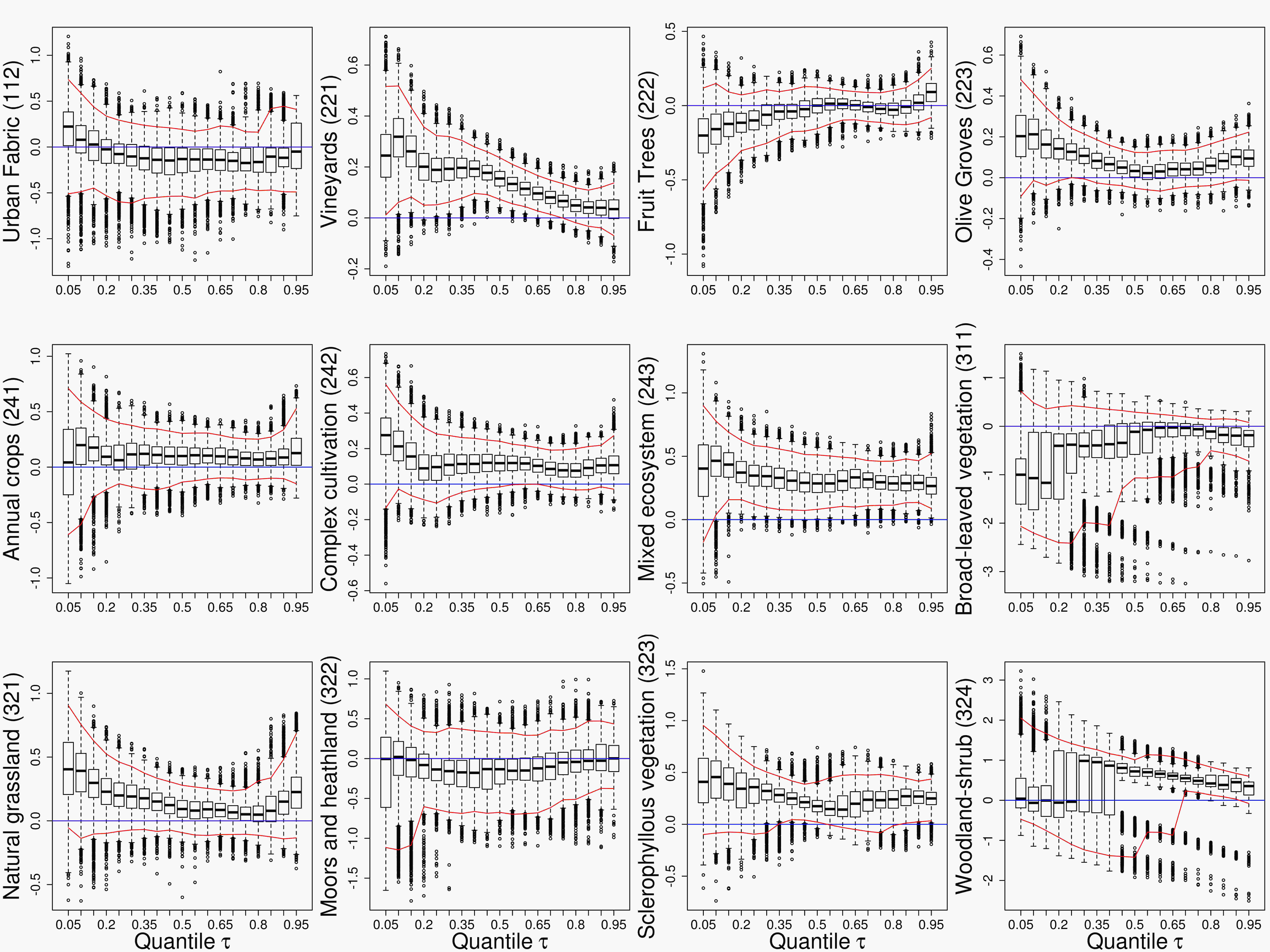}
\caption{Boxplots of estimated beta coefficients for each category of Land Use. These results are based on the final model with 10000 bootstrap replicates, and plotted with respect to the quantile level $\tau=0.05,\ldots,0.95$. The blue line represents 0 (i.e., no effect), while the red curves are $95\%$ pointwise confidence intervals. Numbers between parentheses correspond to the Corine 2000 codes. In particular, Mixed ecosystem corresponds to Land principally occupied by agriculture, with significant areas of natural vegetation (Corine 243).}.
\label{fig:Beta Land Use}
\end{figure}

\begin{figure}[t!]
\centering
\includegraphics[width=0.8\textwidth]{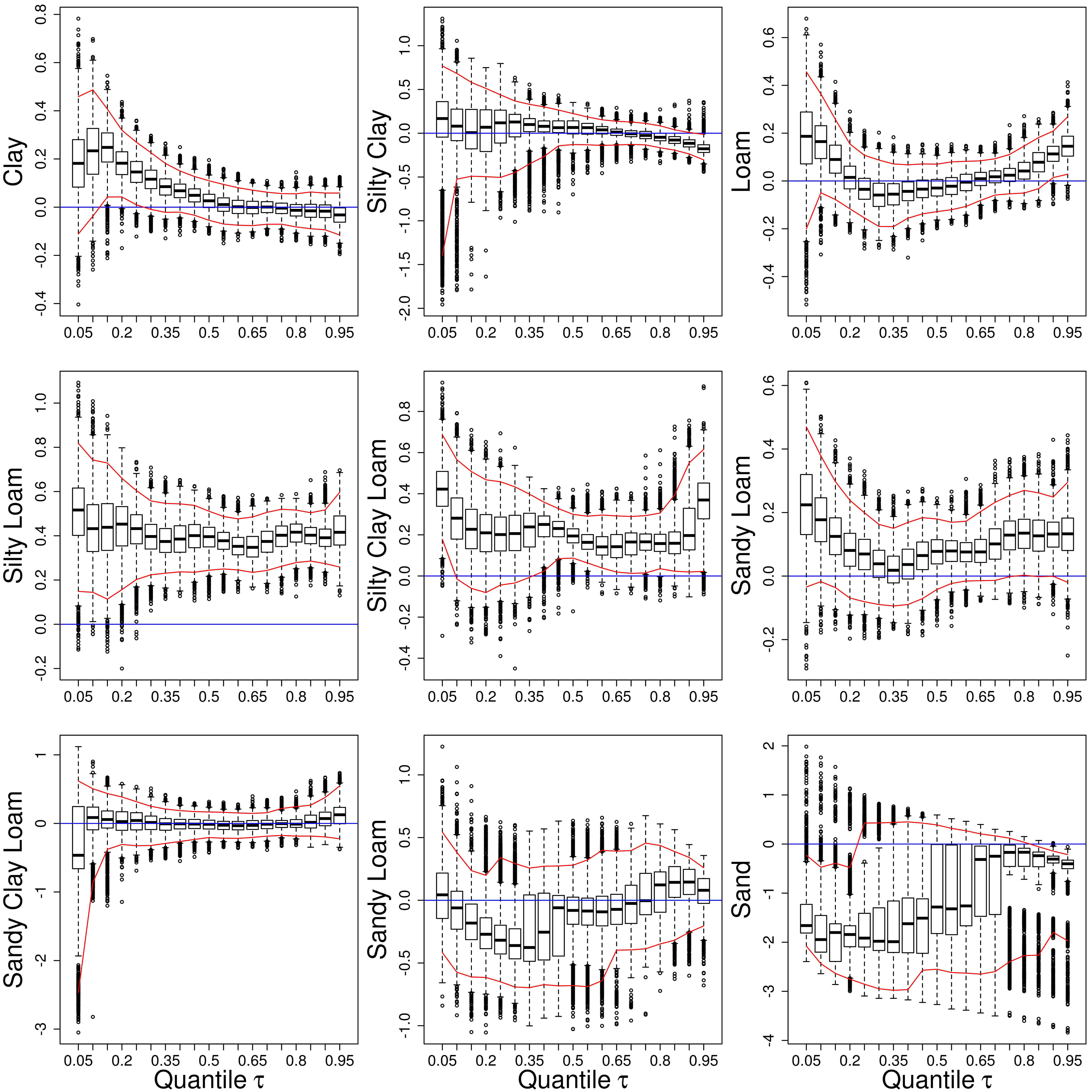}
\caption{Boxplots of estimated beta coefficients for each category of Texture. These results are based on the final model with 10000 bootstrap replicates, and plotted with respect to the quantile level $\tau=0.05,\ldots,0.95$. The blue line represents 0 (i.e., no effect), while the red curves are $95\%$ pointwise confidence intervals.}.
\label{fig:Beta Texture}
\end{figure}

\begin{figure}[t!]
\centering
\includegraphics[width=0.8\textwidth]{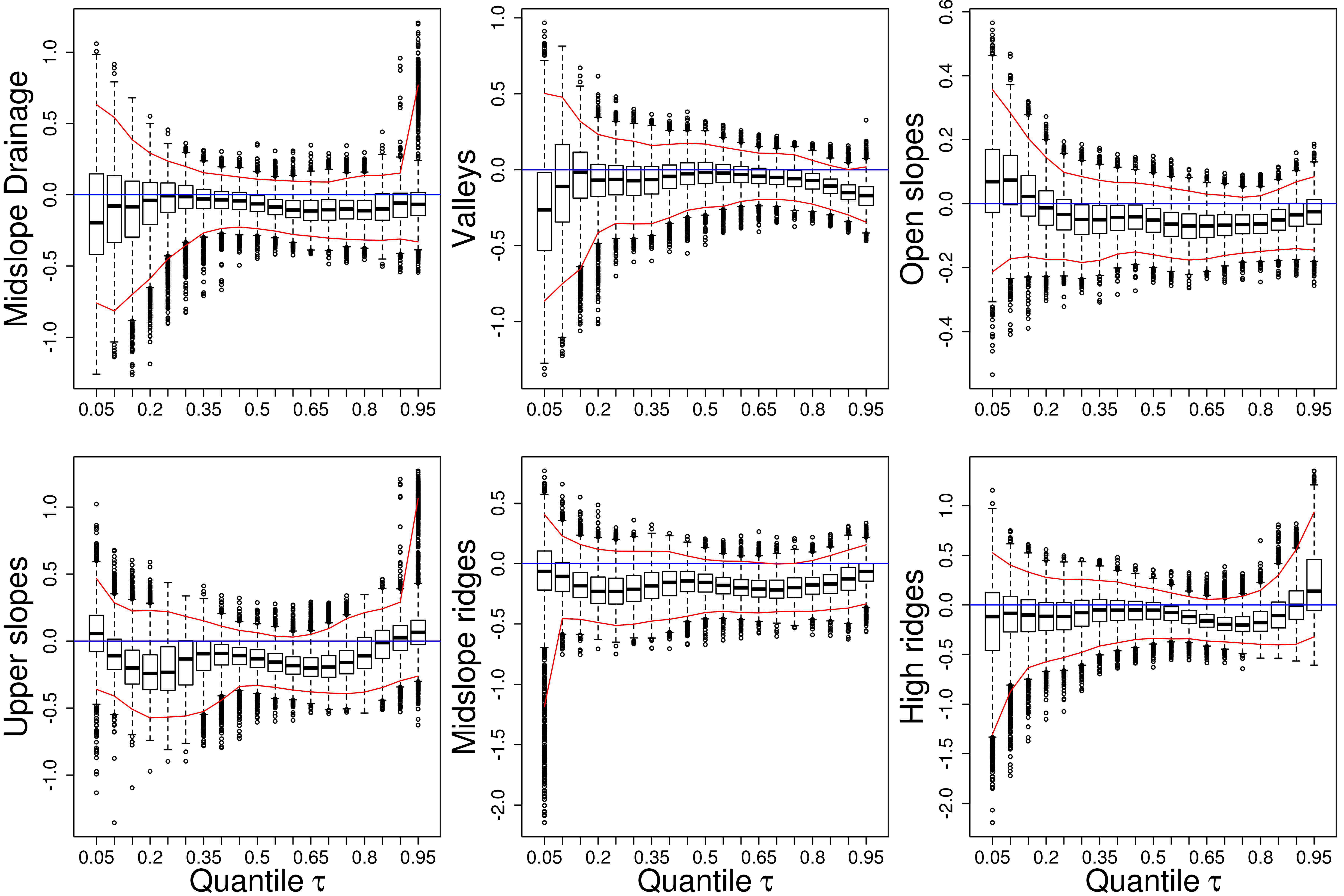}
\caption{Boxplots of estimated beta coefficients for each category of Landform Classification. These results are based on the final model with 10000 bootstrap replicates, and plotted with respect to the quantile level $\tau=0.05,\ldots,0.95$. The blue line represents 0 (i.e., no effect), while the red curves are $95\%$ pointwise confidence intervals.}.
\label{fig:Beta Landforms}
\end{figure}

The uncertainty of estimated beta coefficients (assessed by means of the non-parametric case-resampling bootstrap) is presented in five separate subplots: Figure \ref{fig:Beta simple} presents boxplots of estimated parameters obtained from the 10000 bootstrap replicates for the simple model comprising only three categorical variables; The estimated parameters for the final reference model are summarized in Figures \ref{fig:Beta Continuous}, \ref{fig:Beta Land Use}, \ref{fig:Beta Texture}, and \ref{fig:Beta Landforms}, which correspond to continuous predictors, Land use, Texture and Landform, respectively.

\begin{figure}[t!]
\centering
\includegraphics[width=\textwidth]{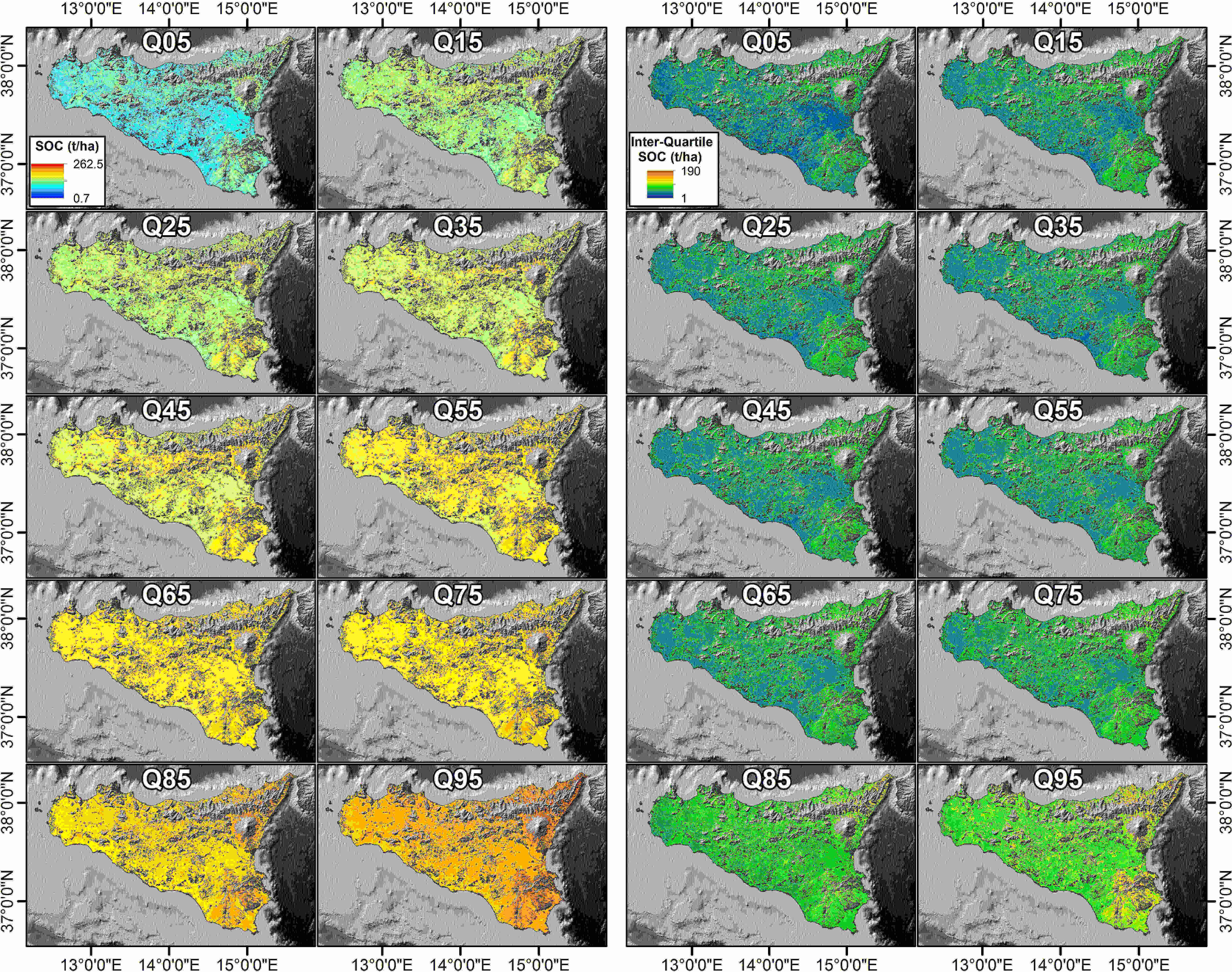}
\caption{Predictive maps (left side) together with their associated variability (right side). The latter is measured as the interquartile range, i.e., the distance between the 75\% and the 25\% quantiles, calculated from the 10000 cross-validated maps. Greyed out regions correspond to no-data zones.}
\label{fig:Prediction}
\end{figure}

\begin{figure}[t!]
\centering
\includegraphics[width=0.95\textwidth]{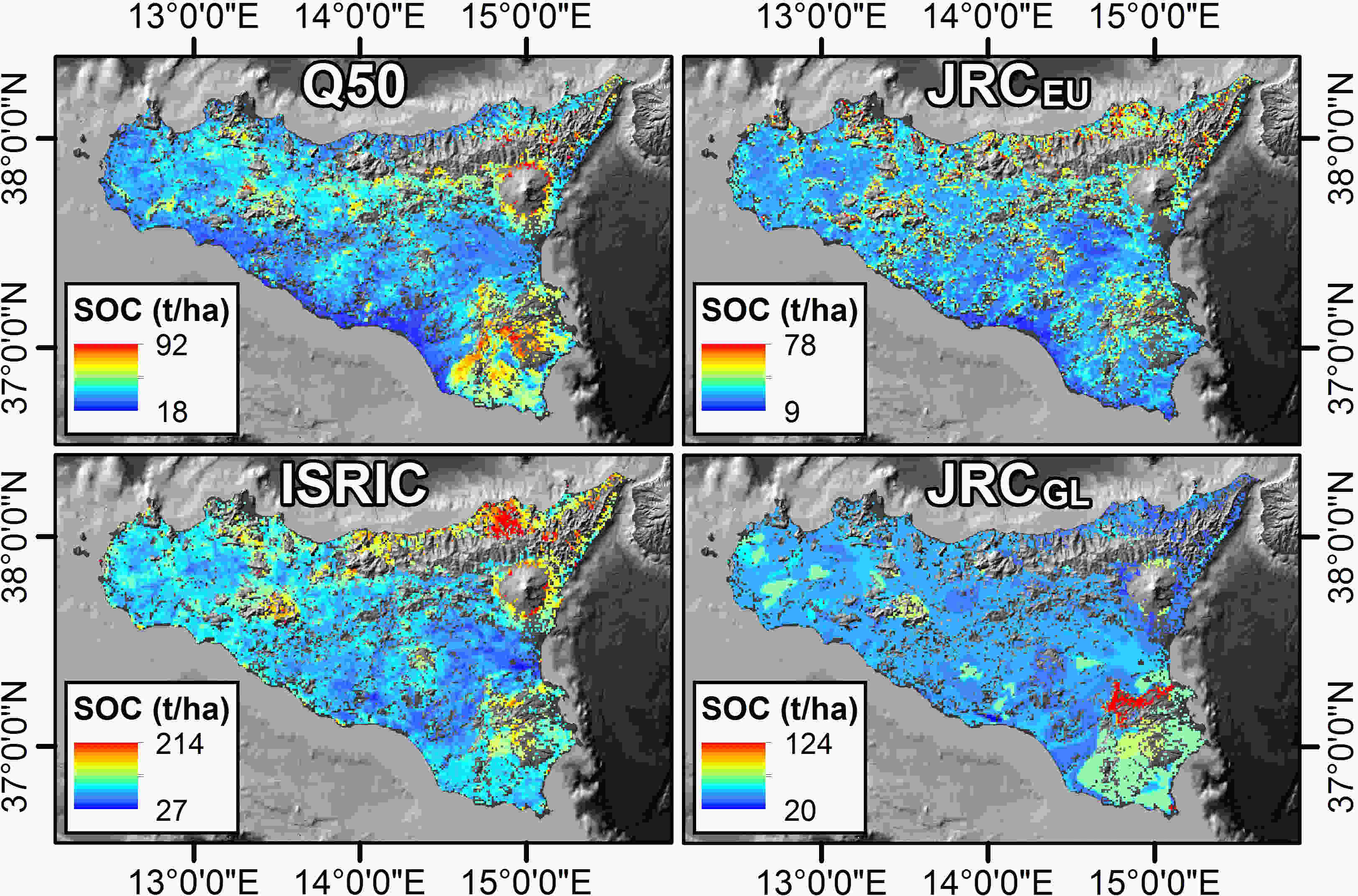}
\caption{Available SOC-stock spatial-predictive maps in Sicily: Q50 corresponds to our median  prediction, ISRIC is the SOC stock map from the International Soil Reference and Information Centre whereas JRC-EU and JRC-GL are the SOC stock benchmarks produced from the Joint Research Centre at the European and Global scale, respectively. Greyed out regions correspond to no-data zones.}
\label{fig:Map Comparison}
\end{figure}

\begin{figure}[t!]
\centering
\includegraphics[width=0.9\textwidth]{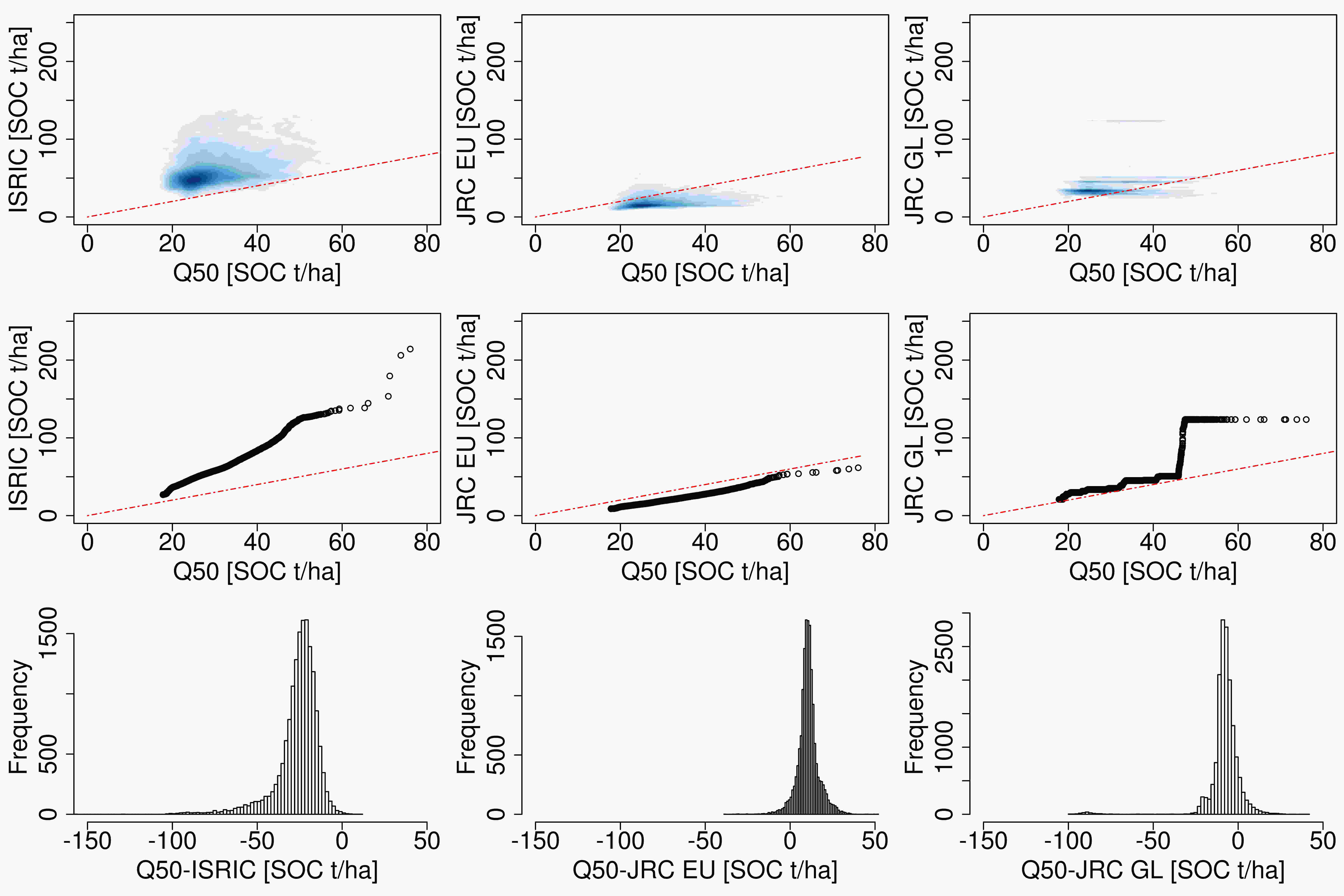}
\caption{Bivariate comparison between median and available benchmark. The first row shows a density scatterplot between our predicted median map and the three available benchmarks in each column. The second row presents the same information compressed in a quantile-quantile plot. The third row summarises the residuals. Red dashed lines correspond to linear fits with regression coefficients equal to 1.}
\label{fig:Benchmarking}
\end{figure}

The spatial prediction and its uncertainty are summarized in Figures \ref{fig:Prediction}, \ref{fig:Map Comparison} and \ref{fig:Benchmarking}. 

The simple model (see Figure \ref{fig:Beta simple}) accounts for the most represented categorical classes in Land use (\textit{Non-irrigated arables}), Texture (\textit{Clay loam}) and Landforms (\textit{Plains}). This model is characterized by a very low variability of the intercept. Non-irrigated arables and Clay loam are negatively associated with SOC, and beta coefficients show a tendency to further decrease at the upper quantiles, especially for the textural class. Plains scarcely influences the SOC stock.

For the final model, some covariates clearly appear dominant, this being shown through high deviations from the blue line corresponding to zero beta coefficient along the quantiles. This particularly occurs for the continuous the covariates \textit{log(Catchment Area)}, \textit{Mean Annual Rainfall} and \textit{Mean Annual Temperature} as shown in Figure \ref{fig:Beta Continuous}. Mean Annual Temperature shows a negative trend at quantiles $\textless$ 0.15. Conversely, \textit{log(Catchment Area)} and \textit{Mean Annual Rainfall} contributions to the prediction are always positive. Other predictors including \textit{Northness}, \textit{Eastness} contribute to SOC stock increase whereas \textit{Slope} reduces it.   

The effects of \textit{Land Use} are reported in Figure \ref{fig:Beta Land Use} where \textit{Vineyards} and \textit{Olive orchards} (Corine Code 221 and 223, respectively) show a positive relationship with organic carbon content and a tendency for beta coefficients to decrease towards the upper quantiles. Conversely, coefficients of \textit{Land principally occupied by agriculture, with significant areas of natural vegetation}, \textit{Natural grassland} and \textit{Sclerophyllous vegetation} are slightly but constantly positive.

The analogous representation for \textit{Texture} is shown in Figure \ref{fig:Beta Texture}. Here, the role of Texture emerges for few textural classes. In particular, \textit{Silty Loam}, \textit{Silty Clay Loam}, and \textit{Sandy} textures appear to be strongly, mildly, and weakly positive, respectively, even across all quantiles. The mean beta coefficient per quantile in \textit{Clay} and \textit{Sand} shows an opposite pattern. On the one side, clay texture yields very high positive beta coefficients at lower quantiles and decreases approximately to zero to the right tail of the SOC distribution. A similar, but less pronounced decrease in the beta coefficients is shown for the \textit{Silty Clay}. Both \textit{Clay} and \textit{Silty clay} present a very low internal variability, especially at the upper quantiles. On the other side, sand texture class produces an increasing beta coefficient across the quantiles, from strongly negative to the left side of the distribution to almost 0 in the right side. However, the variability within each beta coefficient in each quantile for sand is very high, hindering its interpretation.  

Coefficients for \textit{Landform} classes are summarized in Figure \ref{fig:Beta Landforms} (except for \textit{Plains}) where unexpectedly, none of the Landform classes appear to have a clear influence over the SOC in the study area and no pattern across quantiles can be ascertained.  

Predictive maps are shown in Figure \ref{fig:Prediction}. Here, variations in predicted SOC over the study area are evident in the extreme quantiles ($q$ $\leq$ 0.25 and $q$ $\geq$ 0.75) but less pronounced in the central quantiles (0.25 $\textless$ $q$ $\textless$ 0.75). Similarly, the variability (measured as inter-quartile range) shows an increasing trend through quantiles.

The qualitative comparison between the predicted median and those of ISRIC, European and Global JRC benchmarks is shown in  Figure \ref{fig:Map Comparison}. 
Among the available SOC Stock benchmarks, the JRC European map and, to a certain degree, ISRIC map are close to our median map in term of degrees of spatial variability (Figure \ref{fig:Benchmarking}). ISRIC frequently overestimates SOC stock in the study region. In particular, our predicted median and ISRIC maps efficiently capture the pedo-genetic differences but not differences within land use classes. JRC-EU better capture differences within arables, which was far the most represented classes of land use. Finally, JRC-GL captures few spatial differences but, similarly to our predicted median it is the only benchmark capturing the high SOC stocks in the southeastern areas.

The spatial relation between predictive maps is compressed for a numerical-only assessment in Figure \ref{fig:Benchmarking}. 

Here, the reference predicted median is compared to the three benchmarks through i) pixel-by-pixel density plotting, ii) quantile-quantile plot, iii) residuals. Three observations can be made. ISRIC is strongly overestimating the SOC stock compared to our QR-based model only with low-carbon coincident concentrations. The qualitative similarity between the median and the JRC-EU predictions is once more confirmed from a quantitative perspective with a quantile-quantile plot showing a slight but constant underestimation. Ultimately, JRC-GL shows the lowest residuals with respect to the QR reference together with a good agreement up to a concentration of approximately 45 $t/ha$. However, from this threshold to the right tail of the distribution, the two predictions completely diverge one from the other.        

\section{Discussion}
\label{sec:Discussions}

We present a Quantile Regression framework for modeling SOC stock data. This is applied to the semi-arid Sicilian territory located in the middle of Mediterranean Sea. We explore its application evaluating its predictive performances and assess it as a tool to provide a deeper information on predictor effects at different carbon contents. This makes QR a tool to produce reliable soil maps. 

In terms of predictive skills, QR shows comparable results (maximum R$^2$ of 0.49, in Figure \ref{fig:Performances}) to those obtained with Stochastic Gradient Treeboost (R$^2$ of 0.47, \cite{schillaci2017modelling}) using the same dataset. 

Other experiments show equivalent or worse performances. \cite{yigini2016assessment} obtain an $R^2$ coefficient of 0.40 at the European scale with regression-kriging, whereas \cite{meersmans2008multiple} report an $R^2$ coefficient of 0.36 with multiple regression and \cite{nussbaum2014estimating} R$^2$ of 0.35, both at regional scales.   
Quality metric based on the quantile loss highlights a decreased performance near the left and right tails of the SOC stock distribution.  

The simple model intercept (Figure \ref{fig:Beta simple}) shows values bounded between 10 and 130 $t/ha$ which are in line with the original dataset and interestingly these values show a very low variability. This implies that the contribution of \textit{Non-irrigated arables}, \textit{Clay loam}, and, to a lesser extent, \textit{Plains} is very strong.  
Notably, the intercept of the final model (Figures \ref{fig:Beta Continuous}), that also bears the effects of the \textit{Non-irrigated arables}, \textit{Clay loam} and \textit{Plains}, shows values very similar to the simple model but a higher variability. This implies that the greater model complexity due to the inclusion of other predictors (both for continuous and categorical) can produce high ranges of variation in the SOC stock.

\textit{Mean Annual Rainfall} and \textit{log(Catchment Area)} coefficients are constantly positive, confirming the influence of soil moisture on carbon sequestration as reported in several articles \citep[e.g.,][]{saiz2012variation}. Nonetheless, these result partly disagree with \cite{schillaci2017modelling}, that found that found a scarce, but still positive, influence of the untransformed CA on SOC stock of the same area, with a method capable of handling non-gaussian distributed data. This difference points at the need of transforming data even for non-strictly statistical predictive methods. 

In contrast to \textit{Mean Annual Rainfall} and \textit{log(Catchment Area)}, \textit{Mean Annual Temperature} shows negative and slightly varying beta coefficients across the whole SOC distribution. Recent surveys clearly highlight the balance between temperature and rainfall in shaping the background SOC and SOC stocks amount and variations (\citep{davidson2000biogeochemistry,FAO2017,schillaci2017spatio}). 

However, the community still debates whether the temperature should have a positive correlation with SOC stocks \citep[e.g.,][]{conant2011temperature,sierra2015sensitivity}. In the present work, the strong and negative effect of the temperature supports the hypothesis that temperature negatively affects SOC accumulation in agricultural soils of Mediterranean areas even when SOC or rainfall or both are high. This could depend on the erraticness of rainfall and thus water availability that can consist in a low water availability even at high rainfall, which can be lost by runoff \citep{panagos2017towards}. The unclear but apparently low temperature effect and clear and positive rainfall effect at the lowest quantiles also suggests that when SOC is low, management of water availability rather than temperature mitigation should be put in place.

Ultimately, \textit{SL} beta coefficients across quantiles are almost constantly negative confirming the influence of erosion on carbon stocks \citep{olson2016impact}.

From textural classes a general positive trend for mixed granulometries emerges. This is typical for Sicilian soils as sand classes do not have the capacity to fix organic matter while purely clayey soils are extremely variable. A peculiar effect actually characterizes the Clay class with a positive beta coefficient sign from quantile 0.05 to 0.50 aligning to zero values from the median to the 95 percentile. This can be interpreted as a strong clay protective effect for small carbon contents up to a limit where other factors need to interplay in order to further increase the carbon fixation/absorption \citep{badagliacca2017assessment,grimm2008soil,mondal2016impact}. 

Among different uses strong positive relations can be recognized for \textit{Vineyards}, \textit{Olive Groves}, \textit{Land principally occupied by agriculture, with significant areas of natural vegetation}, \textit{Natural Grassland} and \textit{Sclerophyllous vegetation}. \citep{vicente2016soil} report carbon sequestration rates of 0.78 tC ha$^-$$^1$ yr$^-$$^1$ Mediterranean vineyards. Similarly, \cite{farina2017modeling} suggest a potential SOC stock increase of 40.2\% and 13.5\% for vines and olives in similar environments to those considered in this study, respectively. In our work, such a positive effect were found also at the lowest boundary of the SOC distribution. This has a direct implication for land use management when aiming to increase SOC in such a fragile ecosystems compared to arables. In Sicily, arables are mostly winter cereals and grain legumes, which respectively reduce N availability for the microorganisms and have few residues. 

Similarly, the positive effects of \textit{Land principally occupied by agriculture, with significant areas of natural vegetation (Corine 243)} suggest that in-field and in-farm crop and landscape and environmental diversification can also favor SOC accumulation irrespective of the initial SOC levels in semi-arid Mediterranean environments, as also found in continental north-European areas by \cite{kaczynski2017modelling}. Their work cover the time window between 1971 and 2013 during which the authors highlight a marked increase in SOC stock from 2001 coinciding with crop production as a very high yields provided very high input of carbon from crop residues. With respect to \textit{Land principally occupied by agriculture, with significant areas of natural vegetation}, \cite{tian2016soc} conduct a study in China in order to estimate carbon sequestration in different grassland quality condition, which also depend on the diversification of its composition. Their conclusions show that the average sequestration rate was $0.04\cdot 10$$^1$$^2$ $kgC\cdot ha^-1$ and that this rate increases as the grassland quality increases, which also depends on the diversification of its composition. 

As regards the Sclerophyllous vegetation, other studies have highlighted its contribution to SOC even in Mediterranean contexts \citep{munoz2013modelling}.

In terms of soil mapping, 
the four maps (our median and the three benchmarks) agree in depicting higher SOC stock levels around the Etna volcano and generally at the foothills.  This may be interpreted as a result of particle transport where Carbon-rich soil from reliefs are eroded and deposited at the bottom of mountain ranges and/or different geological substrates producing soils with contrasting ability to retain organic C \citep{Costantini2016,mondal2016impact}. A similar agreement is produced in the central portion of the study area but with lower SOC concentrations. Conversely, the southeastern sector is shown to carry high SOC stocks for three maps with the exception of the European JRC, whereas the Global JRC depicts less reasonable patterns and ISRIC overestimates the SOC stock with peaks well above any local measurement. Our SOC stock predictive map shows reasonable values such as JRC and reasonable spatial patterns such as ISRIC. This can clearly be due to a higher resolution because ISRIC, Global and European JRC are continental or global and at such scale the landscape scale is often not represented. Nevertheless, QR was able to reach this level of detail suggesting its use for different datasets and modeling scales.  

\section{Conclusion}
\label{sec:Conclusions}

QR performs similarly to other statistical methods and enables considerations at given sub-domains of the SOC stock distribution. The link between SOC stock amount and the distribution of some Land Use classes (Vineyards, Olive orchards and Mixed ecosystems (Corine 243)) or and presence of Clayey soils was positive and, above all, varying across the SOC distribution. This has direct implication in the management of agriculture at the regional level, since these crops are likely to contemporary increase the gross income of the area and also the ecosystem benefits, such as C sequestration in the soil.

Variables like Vineyards or Clay change significantly through the SOC distribution. This suggest that classical linear regression methods may not recognize this trend and ultimately generate very different SOC values at high or low carbon contents. Furthermore, advantages can be drawn from an agronomic point of view as a better understanding of environmental effects at various SOC concentrations can improve management schemes and allow for sequestration-tailored practices that preserve yield and rentability. This paper shows that Quantile Regression has valid and interesting agronomic applications, as observed in few recent examples \citep{barnwal2013climatic,Yu2016}. To promote its applicability and reproducibility, the R code is made available in the Supplementary Materials.   

\bibliographystyle{plainnat}
\nocite{*}
\bibliography{SOC}
\end{document}

%% file: mylatex.tex
\def\bi{\begin{itemize}}
\def\ei{\end{itemize}}
\def\be{\begin{equation}}
\def\ee{\end{equation}}

\def\pr{\mathrm{pr}}

%% file: packages.tex
\usepackage[english]{babel}
\usepackage{amsfonts}
\usepackage{amssymb}
\usepackage{graphicx}
\usepackage{enumerate}
\usepackage{bm}